\pdfoutput=1

\documentclass{nature_pk}
\usepackage{amsmath} 
\usepackage{url}
\usepackage{graphicx}

\makeatletter
\let\saved@includegraphics\includegraphics
\AtBeginDocument{\let\includegraphics\saved@includegraphics}
\renewenvironment*{figure}{\@float{figure}}{\end@float}
\makeatother

\spacing{1}

\title{\noindent A disk-dominated and clumpy circumgalactic medium of the Milky Way seen in X-ray emission}
\author{
P.~Kaaret$^{1,*}$, D.~Koutroumpa$^{2}$, 
K.D.~Kuntz$^{3,4}$, K.~Jahoda$^{4}$,
J. Bluem$^{1}$, H. Gulick$^{1}$, E.~Hodges-Kluck$^{4}$, 
D. M. LaRocca$^{1}$, R. Ringuette$^{1}$, A. Zajczyk$^{4,5}$
}

\begin{document}

\maketitle

\begin{affiliations}
\item{Department of Physics and Astronomy, University of Iowa, Van Allen Hall, Iowa City, IA 52242, USA} 
\item{LATMOS/IPSL, CNRS, UVSQ Universit\'{e} Paris-Saclay, UPMC Univ. Paris 06, Guyancourt, France} 
\item{The Henry A. Rowland Department of Physics and Astronomy, Johns Hopkins University, Baltimore, MD 21218, USA} 
\item{NASA Goddard Space Flight Center, Greenbelt, MD 20771, USA} 
\item{Center for Space Sciences and Technology, University of Maryland, Baltimore County, 1000 Hilltop Circle, Baltimore, MD 21250} 
\end{affiliations}

\noindent\textbf{The Milky Way galaxy is surrounded by a circumgalactic medium (CGM)\cite{Tumlinson2017} that may play a key role in galaxy evolution as the source of gas for star formation and a repository of metals and energy produced by star formation and nuclear activity\cite{Putman2012}. The CGM may also be a repository for baryons seen in the early universe, but undetected locally\cite{Shull2012}. The CGM has an ionized component at temperatures near $2 \times 10^{6}$~K studied primarily in the soft X-ray band\cite{McCammon2002,Miller2013}. Here we report a survey of the southern Galactic sky with a soft X-ray spectrometer optimized to study diffuse soft X-ray emission\cite{Kaaret2019}. The X-ray emission is best fit with a disc-like model based on the radial profile of the surface density of molecular hydrogen, a tracer of star formation, suggesting that the X-ray emission is predominantly from hot plasma produced via stellar feedback. Strong variations in the X-ray emission on angular scales of $\sim10^{\circ}$ indicate that the CGM is clumpy. Addition of an extended, and possibly massive, halo component is needed to match the halo density inferred from other observations\cite{Weiner1996,Grcevich2009,Gupta2012}.}

\begin{figure*}[t]
\begin{center}
\includegraphics[width=4.5in]{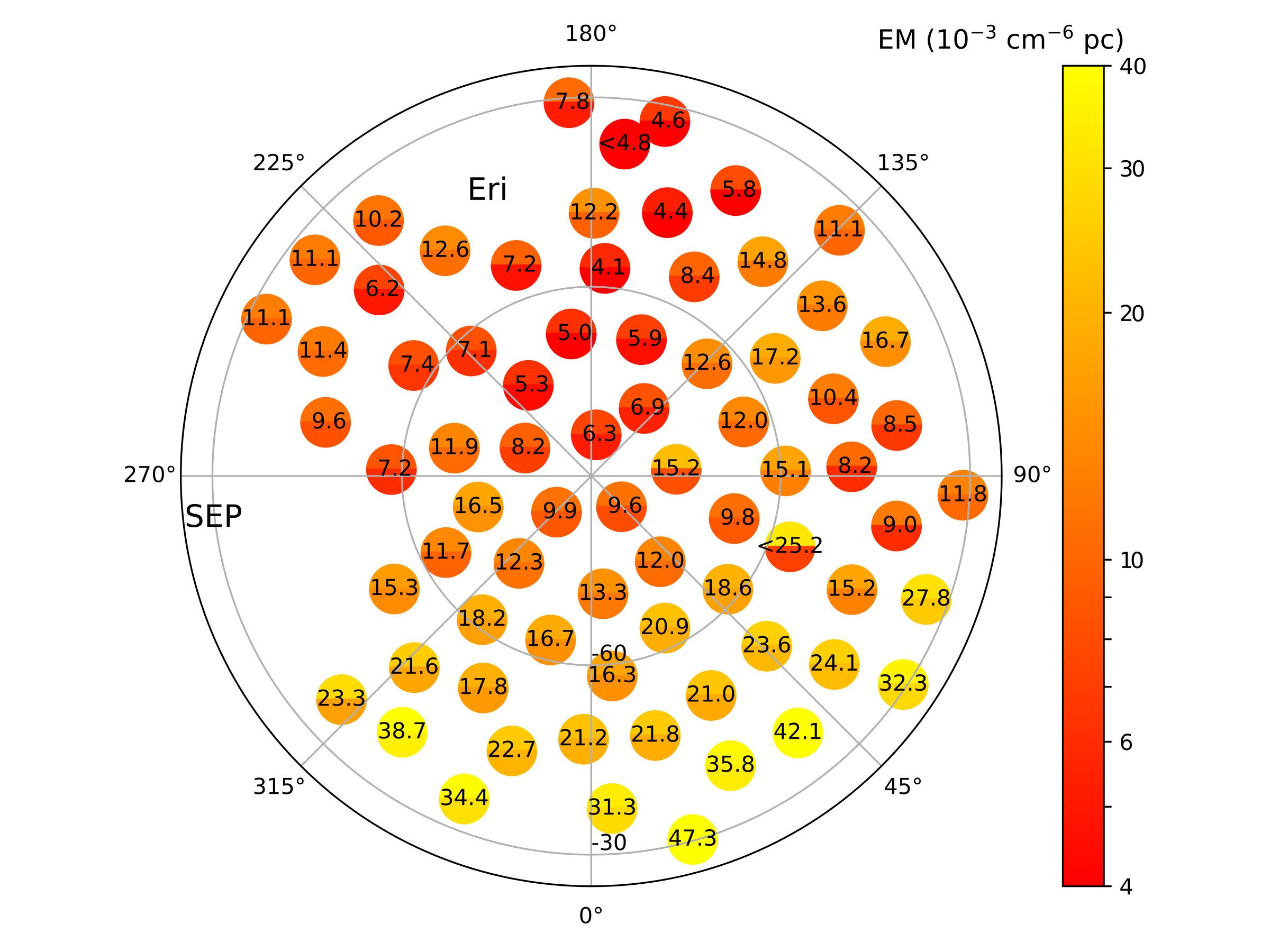} 
\includegraphics[width=4.5in]{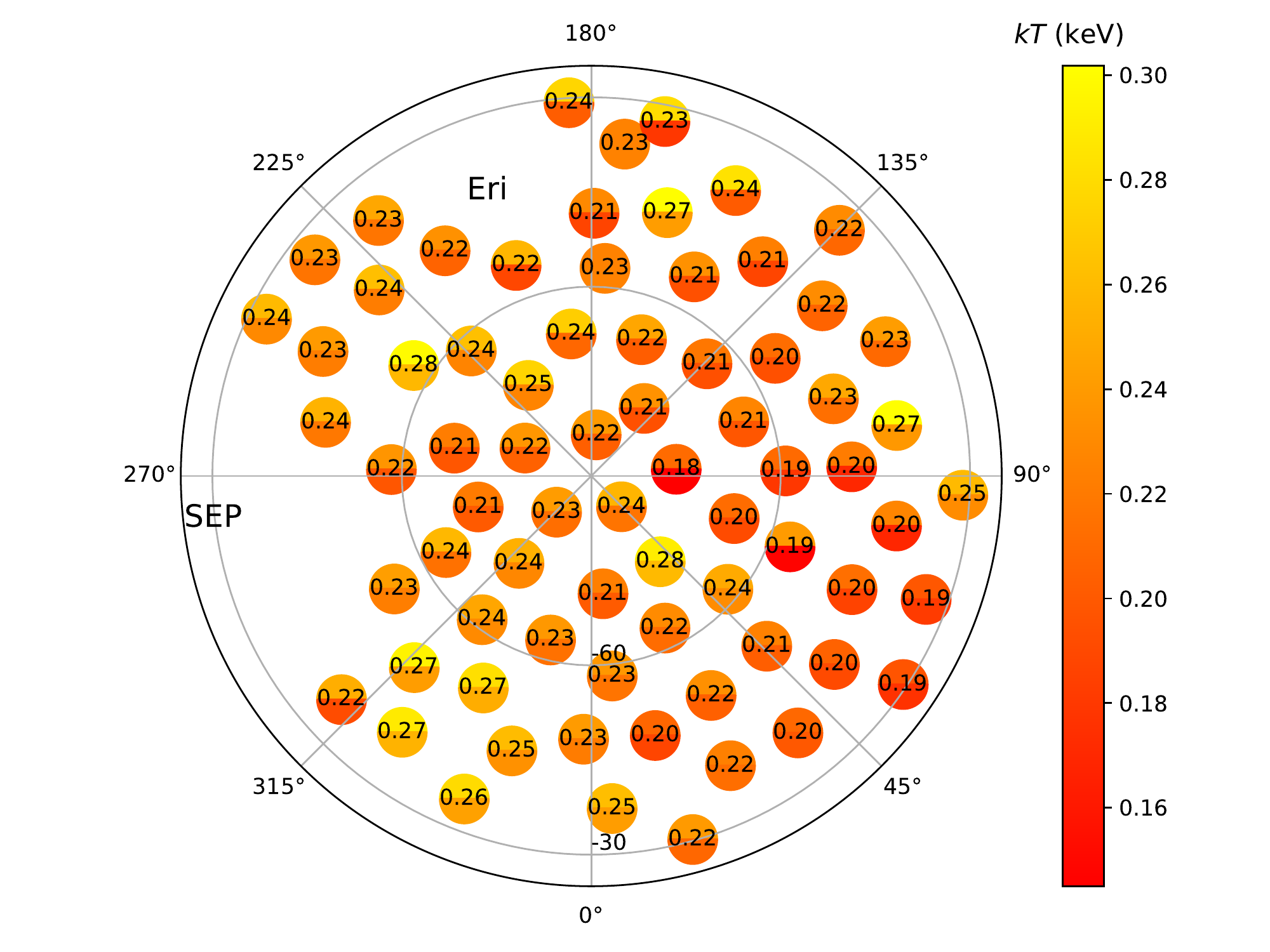}
\end{center}
\caption{Polar plots in Galactic coordinates of the emission measure (EM) and temperature ($kT$) of the HaloSat fields with Galactic latitudes $b < -30^{\circ}$. The top panel shows the emission measure (EM) in units of $\rm 10^{-3} \, cm^{-6} \, pc$ and the bottom shows the temperature ($kT$) in units of keV of the thermal plasma component in the X-ray spectral fits representing the halo emission. The color of the top half of each circle indicates the measured value plus the 90\% confidence error while the bottom half indicates value minus the error. The Galactic south pole is at the centre of each plot. Two fields within the Eridanus enhancement\cite{Snowden1995}, indicated on the EM map as `Eri', were removed. The Sun angle cut removes fields near the South Ecliptic Pole shown as `SEP'. Some overlapping fields were removed for clarity.}
\label{em_kT_maps}
\end{figure*}

Figure~\ref{em_kT_maps} maps the X-ray emission from the southern halo (Galactic latitudes $b < -30^{\circ}$) of the Milky Way as modeled by thermal emission from a collisionally ionized plasma characterized by the temperature of the gas ($kT$) and the emission measure (EM), the integral along the line of sight of the product of the electron and ion densities\cite{Miller2013}. The data were obtained with HaloSat, which is a CubeSat-based X-ray observatory\cite{Kaaret2019}, optimized to study diffuse X-ray emission with a field of view of $10^{\circ}$ diameter full response tapering to zero at $14^{\circ}$. X-ray emission from solar-wind charge exchange (SWCX) interactions in the heliosphere\cite{dimitra_hswcx}, the Local Hot Bubble (LHB) of gas within $\sim 200$~pc of the Sun\cite{Liu2017}, the cosmic X-ray background\cite{Capuletti2017}, and the particle-induced instrumental background were modelled and included in the spectral fitting (see Methods). We fix the metallicity, the fraction of heavy metals in the plasma, to 0.3 of the solar value\cite{Miller2015}. We estimate the absorption column density for each field from optical reddening\cite{Zhu2017}.


\clearpage

The temperatures ($kT$) have a median of 0.225~keV and a standard deviation of 0.023~keV, about 50\% larger than the measurement uncertainties. The temperatures are similar to those found in previous work; the median is slightly lower than that obtained in\cite{Nakashima2018} and slightly higher than that in\cite{Henley2013}. The differences are likely due to how the LHB is modelled\cite{Nakashima2018}. Because the peak of the LHB lies below the low-energy spectral cutoff at 0.4~keV of HaloSat (the same cutoff as the other analyses), we fix the LHB temperature to the average value and fix the LHB EM using the SWCX-corrected ROSAT All Sky Survey sensitive in the 1/4~keV band near the peak of the LHB emission\cite{Liu2017}.

Previous work\cite{Henley2012,Nakashima2018} shows strong variations in the halo EM or oxygen line flux for closely aligned pointing directions, likely caused by foreground SWCX emission. Reference\cite{Henley2012} characterizes variations in a set of fluxes measured within $0.1^{\circ}$ of a given line of sight as the fractional excess over the minimum flux in the set. Their median excess for O{\sc vii} line emission is 0.45 after filtering based on the solar wind with 22\% of the observations having an excess greater than 1. For repeated HaloSat observations, we find a median EM excess of 0.21 and 6\% of observations with an excess greater than 1, suggesting a significantly lower level of SWCX contamination. The observations in our halo analysis were obtained at Sun angles $\ge 110^{\circ}$ to avoid lines of sight through the magnetosheath to minimize magnetopheric SWCX contamination and to reduce the temporal variation in the heliospheric SWCX\cite{kuntz_swcx}. This was not possible for\cite{Henley2013,Nakashima2018} due to the Sun angle constraints of the observatories used. Also, the observations in\cite{Henley2013} were spread over 10 years with more than half during periods of high solar activity with $\rm O^{7+}$ fluxes elevated by a factor $\sim5$ compared to the period of low solar activity during which the HaloSat observations were acquired. Furthermore, we carefully modelled and removed the heliospheric SWCX emission for each HaloSat observation based on contemporaneous solar wind measurements (see Methods). All of these factors help reduce contamination due to foreground emission in the HaloSat data relative to previous studies.

We evaluated the statistical error from spectral fitting versus the fluctuations in the EM and $kT$ for repeated observations of individual targets by calculating the deviation of EM and $kT$ from the weighted average for the target divided by the statistical error. Applying a correction for the finite sample size, the standard deviations are 1.4$\times$ the statistical fluctuations for EM and $1.3\times$ for $kT$. The fluctuations are likely caused by changing and unmodelled foreground SWCX emission or errors in modelling the time-dependent instrumental background. 

\begin{figure}[t]
\begin{center}
\includegraphics[width=3.25in]{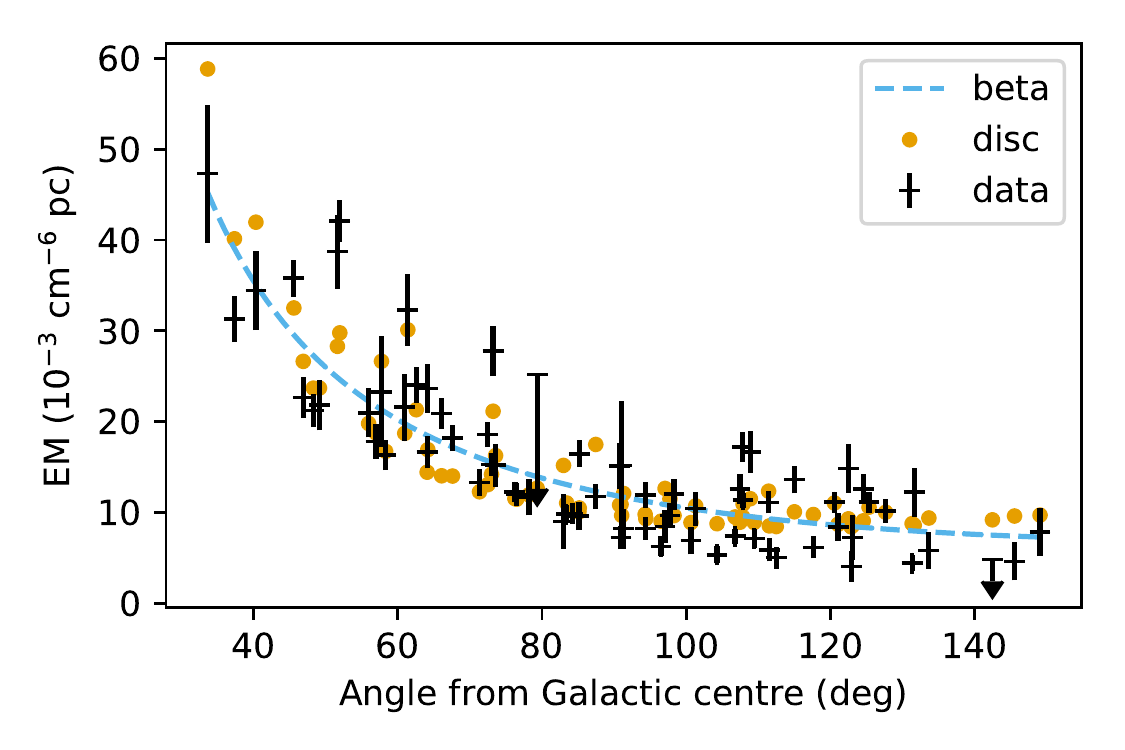}
\includegraphics[width=3.25in]{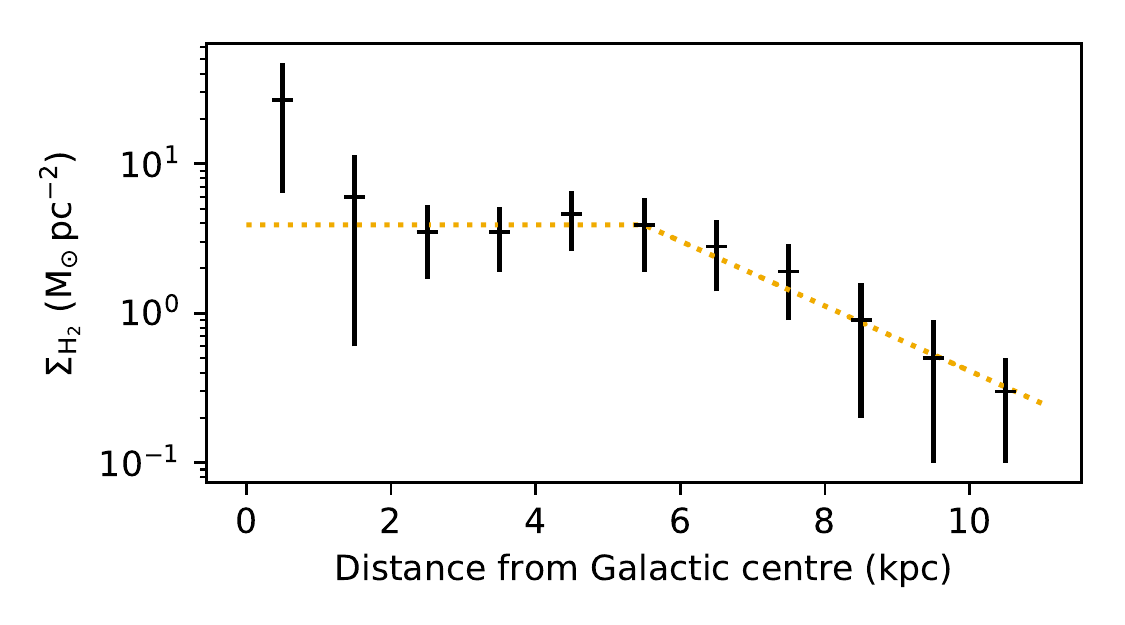}
\end{center}
\caption{\textbf{Model fits to emission measure (EM).} 
\textbf{Top: EM versus angular distance ($\theta$) from the Galactic centre.} The black crosses are the HaloSat measurements. The errors indicate the statistical 90\% confidence intervals (see text for further discussion) and non-detections are plotted as $2\sigma$ upper limits. The dashed blue line represents the best fit modified $\beta$ model. The orange dots represent the best fit disc-like model calculated for each field. The disc-like emission is not a unique function of $\theta$ thus this model does not appear as a continuous curve.
\textbf{Bottom: Empirical molecular gas surface density profile.} The black crosses show the surface density of molecular gas ($\Sigma_{\rm H_2}$) versus distance ($R$) from the Galactic centre\cite{Nakanishi2006}. The yellow dotted line shows a simple empirical model: 
$ \Sigma_{\rm H_2}(R) = (3.9 \, M_{\odot} {\rm \, pc^{-2}}) 
                        e^{-(R - 5.5 {\rm \, kpc})/{\rm 2 \, kpc}}$
for $R > \rm 5.5 \, kpc$ and $3.9 \, M_{\odot} {\rm \, pc^{-2}}$ for smaller $R$.}
\label{em_theta}
\end{figure}

The EM values vary over a wide range, generally increasing closer to the Galactic centre, see Fig.~\ref{em_theta}. There are variations in EM between nearby fields that are significantly larger than the measurement errors. In model fitting and in Fig.~\ref{em_kT_maps}, we use the 90\% confidence statistical errors from the spectral fitting, which overestimate the measured root-mean-squared variations of the EM in repeated observations of individual targets by 10\%.

We compared the EM data with model density distributions. We estimated the EM for each line of sight for each model by integrating 260~kpc along the line of sight in $10^{4}$ steps. To account for intrinsic EM variations, a `patchiness parameter' ($\sigma_p$) is combined with the measurement uncertainty ($\sigma_m$) to obtain the uncertainty adopted in the model fitting\cite{Savage2009}, $\sigma_f^2= \sigma_p^2+ \sigma_m^2$. The patchiness parameter was fixed to $\sigma_p = 3.4 \times 10^{-3} \rm \, cm^{-6} \, pc$ to give a reduced $\chi^2 \approx 1$ for our final model fit. Consistent results were obtained using a robust fitting method with a Huber loss function (see Methods). The statistical errors for the density model fits are quoted at the $1\sigma$ confidence level.

We first examine two density distributions used in the literature, a disc-like morphology and an extended spherical distribution. In the spherically-symmetric, modified $\beta$ model\cite{Miller2013}, the density of ionized matter given by the expression $n(r)= n_c (r/r_c )^{-3\beta}$ where $r$ is the distance from the Galactic centre, $\beta$ sets the slope of the density profile, $r_c$ is a characteristic radius, and $n_c$ is the density at $r_c$. Both $n_c$ and $r_c$ contribute only to the model normalization, so we set $r_c = 2.4$~kpc with no loss of generality\cite{Nakashima2018}. In the exponential disc-like model, the density is given by $n(R,z)= n_0 e^{-R/R_0} e^{-z/z_0}$, where $R$ is the distance from the Galactic centre projected onto Galactic plane, $R_0$ is the exponential scale length, $z$ is the height above Galactic plane, $z_0$ is the scale height, and $n_0$ is the density at the Galactic centre. The best fit parameters for the modified $\beta$ model are $n_c = 0.0044 \pm 0.0004 \rm \, cm^{-3}$ and $\beta = 0.39 \pm 0.02$ with $\chi^2/\rm DoF = 98.2/71$. The best fit parameters for the disc-like model are $n_0 = 0.012 \pm 0.007 \rm \, cm^{-3}$, $R_0 = 5.4 \pm 1.5$~kpc, and $z_0 = 2.8 \pm 1.0$~kpc with $\chi^2/\rm DoF = 87.7/70$. The disc-like model is preferred to the $\beta$-model with an F-test probability of 0.005.

\begin{figure}[t]
\begin{center}
\includegraphics[width=3.25in]{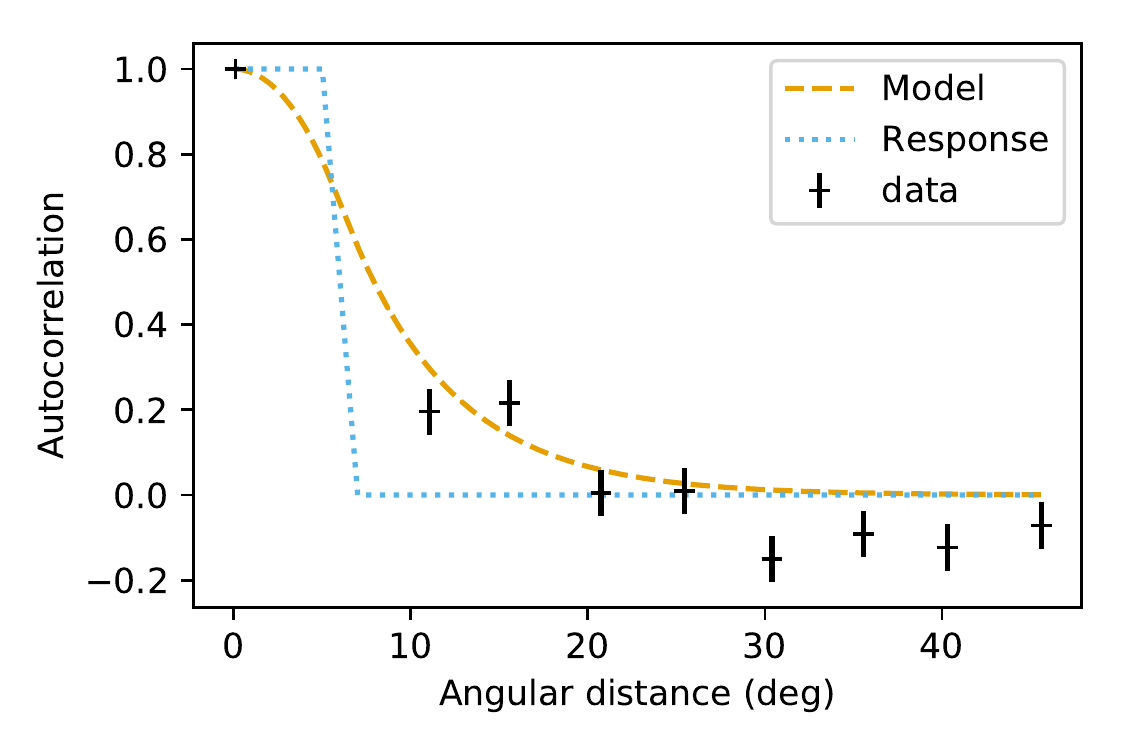}
\end{center}
\caption{Autocorrelation of the deviations from the best fit empirical disc-like model. The black crosses are the data with errors indicating 1-$\sigma$ confidence intervals. The blue dotted curve is the HaloSat angular response. The yellow dashed curve shows the function $w_0(\phi) = e^{-\phi/\phi_a}$ with $\phi_a = 6^{\circ}$ convolved with the HaloSat angular response.}
\label{autocorrelation}
\end{figure}

A disc-like morphology can be produced by stellar feedback in which star-forming regions spawned from molecular gas near the Galactic disc heat and ionize gas that rises vertically and produces the observed X-ray emission\cite{Norman1989}. The density of the ionized gas at high Galactic latitudes might then be expected to scale with the surface density of molecular gas in the disc\cite{Schmidt1959}. The radial profile of the molecular gas surface density in the Milky Way is not well described by an exponential function\cite{Nakanishi2006}. To replace the exponential profile used in the disc-like model above, we derived a simple empirical expression that adequately matches the measured molecular profile, see Fig.~\ref{em_theta}. The density profile is $n(R,z)= n_0 \, \Sigma(R) \, e^{-z/z_0}$, where $\Sigma(R) = 1$ for $R < 5.5$~kpc and $\Sigma(R) = \exp(-(R - 5.5)/2)$ for $R > 5.5$~kpc with $R$ in units of kpc. Use of the empirical radial profile improves the fit with $\chi^2/\rm DoF = 75.8/71$, strengthening the interpretation of the emission as due to stellar feedback. The best fit parameters for the empirical disc-like model are $n_0 = 0.0144 \pm 0.0007 \rm \, cm^{-3}$ and $z_0 = 1.09 \pm 0.08$~kpc. We characterize the angular scale of the fluctuations of the EM around the smooth variations using the autocorrelation function of the deviations of the data from the best fit empirical disc-like model, see Fig.~\ref{autocorrelation}. Accounting for the angular response of the HaloSat detectors, the intrinsic autocorrelation is adequately modeled by the form $w_0(\phi) = e^{-\phi/\phi_a}$ with $\phi_a \approx 6^{\circ}$.

\begin{figure}[t]
\begin{center}
\includegraphics[width=3.25in]{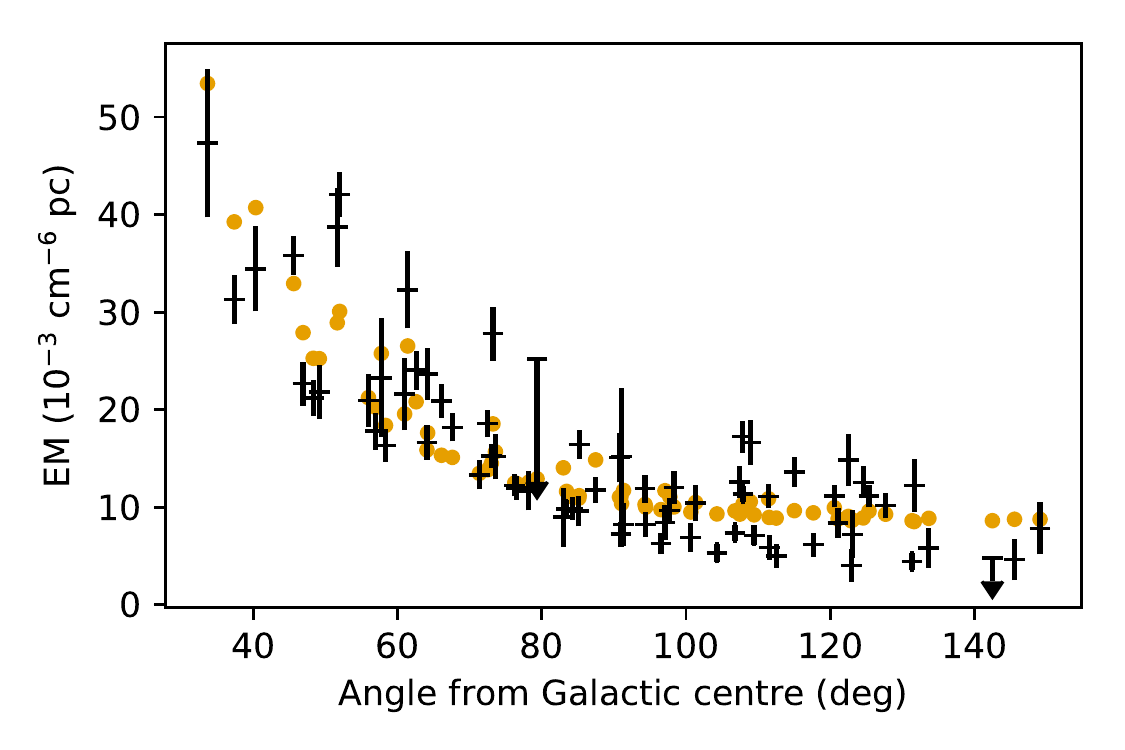}
\includegraphics[width=3.25in]{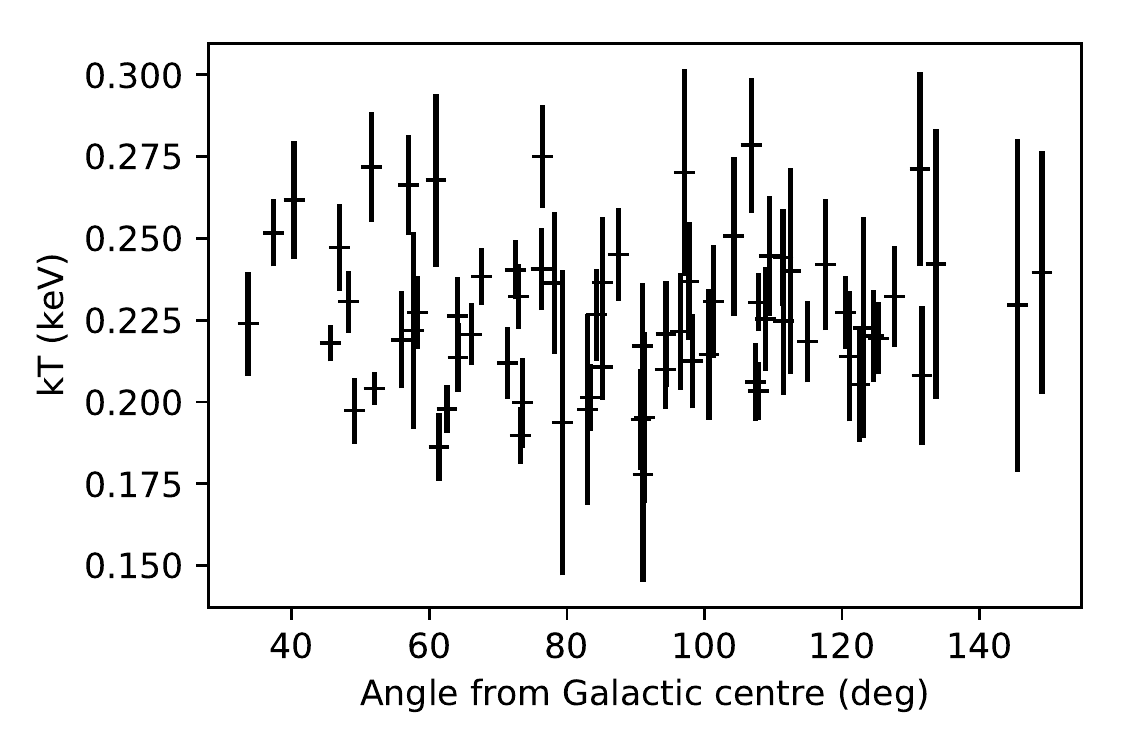}
\end{center}
\caption{\textbf{Model fits to emission measure (EM).} 
\textbf{Top: EM versus angular distance ($\theta$) from the Galactic centre.} The black crosses are the HaloSat measurements as in Fig~2. The orange dots represent the best fit disc-like model consisting of the sum of disc-like and extended halo components described in the text.
\textbf{Bottom: $kT$ versus angular distance ($\theta$) from the Galactic centre.} The black crosses are the HaloSat measurements. The errors in both panels indicate 90\% confidence intervals.}
\label{em_theta_combined}
\end{figure}

The best fit empirical disc-like model significantly underpredicts the O{\sc VII} surface densities measured via X-ray absorption\cite{Gupta2012}, suggesting the presence of an extended low-density halo.  H$\alpha$ emission from the Magellanic Stream\cite{Weiner1996} and ram pressure stripping of satellite galaxies\cite{Grcevich2009,Salem2015} imply $n_H \sim10^{-4} \rm \, cm^{-3}$ at radii $\sim$50~kpc. We fit the HaloSat EM data using the sum of the empirical disc-like model and an extended halo modelled as an adiabatic gas with a polytropic index of 5/3 in hydrostatic equilibrium in the gravitational potential of a dark matter halo with a Navarro, Frenk, and White profile\cite{Maller2004}, see Fig.~\ref{em_theta_combined}. We adopt the gas density profile parameters from\cite{Fang2013}, but fit for the gas density at the virial radius ($\rho_V$). This led to a modest improvement over the empirical disk model (F-test probability of 0.049), but represents a highly significant improvement relative to the $\beta$-model (F-test probability of $3 \times 10^{-6}$). The scale height, $z_0 = 1.60 \pm 0.34$~kpc, is marginally larger than the value obtained for the empirical disc-like model alone, while the central density, $n_0 = 0.0081 \pm 0.0022 \rm \, cm^{-3}$, is slightly reduced. The best fit density at the virial radius is $\rho_V = (4.8 \pm 1.0) \times 10^{-5} \rm \, cm^{-3}$. 

\clearpage

Uncertainty in the absorption column versus optical reddening relation is a potential source of systematic error, so we examined the effect of using a different relation\cite{Rachford2002}. This produces a shift in the EM of each field smaller than the statistical error, but a small, systematic decrease in the EM. The best fit parameters for density model become $z_0 = 2.19 \pm 0.54$~kpc, $n_0 = 0.0053 \pm 0.0015 \rm \, cm^{-3}$, and $\rho_V = (5.4 \pm 0.6) \times 10^{-5} \rm \, cm^{-3}$, consistent within the uncertainties. Also, we have assumed a sub-solar metallicity, $Z = 0.3 Z_{\odot}$. Using solar metallicity, as may be more appropriate for the disk-like component, rescales the EM by a factor of 0.3 reducing the inferred mass by the same factor, but has no significant affect on the other parameters for the spectral fits or the density model fits.

The halo density at 50~kpc is $(1.2-1.9) \times 10^{-4} \rm \, cm^{-3}$ in agreement with the Magellanic Stream and dwarf galaxy results. The predicted O{\sc vii} column densities are $\log(N_{\rm O VII}/\rm cm^{-2}) = 15.7-16.1$ for lines of sight with $|b| > 30^{\circ}$ calculated assuming an O{\sc vii} ion fraction of 0.5 appropriate for a temperature of $2 \times 10^6$~K, a metallicity of 0.3 solar, and integrating to the virial radius taken to be at 260~kpc\cite{Fang2013}. These are reasonably consistent with measured values\cite{Gupta2012,Miller2013,Hodges2016}. The mass in the disc-like component is $2 \times 10^{7} \, M_{\odot}$. Integrating the mass profile to the virial radius gives a halo mass of $(5.5 - 8.6) \times 10^{10} \, M_{\odot}$. This would be increased for lower halo metallicity. Depending on the radial density profile and metallicity, the extended halo may contain sufficient mass to account for the Milky Way's `missing' baryons.


We find that the soft X-ray emission from the Milky Way's circumgalactic medium is best described with a disc-like model, but an extended halo is necessary to satisfy other observational constraints but produces relatively little X-ray emission. The variations on angular scales $\sim10^{\circ}$ in X-ray emission are naturally produced by spatial variations in the intensity of the star formation. The midplane density of the disc-like component at the solar circle is $(1.3 - 2.7) \times 10^{-3} \rm \, cm^{-3}$, which is similar to values found in magnetohydrodynamic simulations of stellar feedback at the solar circle\cite{Hill2012}. The exponential scale height of 1.6-2.5~kpc is similar to values measured for the soft X-ray profiles of edge-on spiral galaxies\cite{Putman2012}, but smaller than the scale height of transition temperature gas\cite{Savage2009}. For a scale height of 2~kpc at $|b| = 30^{\circ}$, the $\sim 6^{\circ}$ angular scale of the EM fluctuations corresponds to a spatial scale of $\sim 400$~pc, comparable to the size of superbubbles found within the Milky Way. The magnitude of the fluctuations, indicated by $\sigma_p$ which is $\sim$30\% of the median EM, is related to variations in the local surface density of star-formation in the disc. The X-ray emission is dominated by the high density regions near the disc because emission is proportional to density squared, but the disc-like component contains little mass due to the small scale height. The halo dominates the X-ray absorption because of the long path length and the fact that absorption is linearly proportional to density.






\begin{addendum}
 
\item[Acknowledgements] We acknowledge support from NASA grant NNX15AU57G and the Iowa Space Grant consortium. D.K.'s modeling work was supported by CNES and performed with the High Performance Computer and Visualisation platform (HPCaVe) hosted by UPMC-Sorbonne Universit\'es. We thank Jim Raines and the SWICS team for providing the solar wind data. We thank the reviewers for improving the paper.

\item[Author Contributions] P.K. did the X-ray data analysis and wrote the text; D.K did the heliospheric SWCX modelling; D.M.L. wrote the code to calculate the absorption column densities and the LHB emission measures; D.K., K.D.K., E.H.-K., K.J., D.M.L., A.Z., R.R, J.B. and H.G. read and commented on the manuscript.

\item[Competing Interests] The authors declare that they have no competing financial interests.

\item[Correspondence and requests for materials] should be addressed to P.K.

\end{addendum}

\clearpage

\noindent{\sffamily\bfseries\large Methods}

\noindent{\bf Observations, data reduction, and spectral modelling.} All HaloSat data for observation fields with Galactic latitude $b < -30^{\circ}$ were processed from spacecraft telemetry into `unfiltered files' that record the arrival time and energy for each X-ray and background event along with housekeeping information regarding the instrument temperature and other parameters\cite{Kaaret2019,LaRocca2020}. These data were filtered using the event counting rates binned in 64~s intervals, requiring the rate in the `hard' band (3-7~keV) to be below 0.12~c/s and the rate in the `Very Large Event' band (above 7~keV) to be below 0.75~c/s. Targets with an average exposure per detector of 2000~s or greater were retained for analysis.

The three energy spectra, one for each of the three X-ray detectors on HaloSat, obtained for each observation were analysed using the Xspec software\cite{Arnaud_xspec}. The pulse height for each X-ray event was converted to energy using a temperature-dependent calibration and the energy redistribution matrix and effective area were calculated from calibration data obtained on the ground\cite{Zajczyk2020} and verified on orbit\cite{Kaaret2019}. The spectra for each target were analyzed with a model consisting of an optically-thin, thermal plasma model (the astrophysical plasma emission code\cite{apec_model} or `APEC', version v3.0.9 with 201 tabulated temperatures) for the Galactic halo emission with the elemental abundances set to 0.3 of the solar value and the temperature and normalization as free parameters, a second APEC model for emission from the Local Hot Bubble with fixed parameters calculated from\cite{Liu2017} based on the target coordinates, two Gaussian components to model oxygen line emission from heliospheric SWCX with fixed line fluxes calculated via modelling described further below based on the target coordinates and observation times, a power-law model for the cosmic X-ray background (CXB) component with fixed parameters calculated from\cite{Capuletti2017}. The parameters of these components were forced to be the same for each detector. For fields where the halo temperature could not be accurately determined, it was set to the median of 0.225~keV. We used a power-law model for the particle-induced instrumental background with the normalization for each detector as a free parameter and the photon index for each detector calculated using the procedure described further below.  We used Wilms abundances\cite{tbabs_ref} and Verner cross-sections\cite{Verner1996}. An example of the HaloSat spectra for one target is shown in Extended Data Fig.~1.



The CXB and halo emission were subjected to interstellar absorption modelled with TBabs model\cite{tbabs_ref}, version 2.3. To estimate the absorption column density, we used maps of optical extinction, E(B-V), measured with the Planck satellite based on the dust radiance with point sources removed\cite{Planck_extinction,dust_python}. Total hydrogen column densities were calculated from the E(B-V) values over a grid of points in each HaloSat field via a relation appropriate for the TBabs model\cite{Zhu2017}. The curves of absorption versus energy were then weighted according the relative response within the HaloSat field and an equivalent total absorbing column density was found by fitting the weighted-average absorption curve. This value was fixed in the spectral fitting.

\noindent{\bf Instrumental background model.} A power-law was used to model the instrumental background for each detector with the photon index calculated from the count rate in the hard (3-7 keV) band for the observation. A linear relation between photon index and hard band count rate was found by performing spectral fits with the instrumental background photon index as a free parameter (with all other parameter as described in the previous section) for targets with an average exposure per detector of at least 8000~s. The targets towards the South Ecliptic Pole (including one target centred on the Large Magellanic Clouds) and the Eridanus enhancement were excluded. The linear fit parameters are given in Extended Data Fig.~2.

We tested the background model by filtering with a cut on the hard band rate of 0.24~c/s, twice that used in our primary analysis. We repeated the full analysis and examined the difference in the fitted parameters for the APEC model for the halo emission for each target. The median of the absolute value of the difference in EM divided by the statistical error is 0.32 and the mean difference in EM is $0.5 \times 10^{-3} \rm \, cm^{-6} \, pc$. The median of the absolute value of the difference in $kT$ divided by the statistical error is 0.13 and the mean difference in $kT$ is $-0.0016$~keV. Thus, the uncertainty in EM and $kT$ due to the instrumental background is smaller than the statistical uncertainties, however, there may be a small systematic shift of the EM on the order of 4\% of the median EM.


\noindent{\bf Estimation of oxygen flux due to solar wind in the heliosphere.} Charge exchange by energetic ions in the solar wind with neutral atoms, predominantly H and He, in the heliosphere can produce excited ions that decay via emission of photons indistinguishable from line emission at the same energy from the Galactic halo\cite{Cox1998}. The strongest heliospheric lines of interest for HaloSat are the O{\sc vii} triplet at 560-574~eV produced by O$^{7+}$ ions in the solar wind charge exchanging to become O$^{6+}$ ions in an excited state and the O{\sc viii} line at 653.1~eV produced by O$^{8+}$ ions.  The line intensity along the line of sight (LOS) in $\rm photons \, cm^{-2} \, s^{-1} \, sr^{-1}$ (Line Units) is calculated from the equation:

\begin{equation}
\label{eq:hswcx_line}
I_\gamma = \frac{1}{4 \pi} \sum_{j=1}^{N} F_{\rm Xq+}(R_j)
  \left[
  N_{\rm H}(\lambda_j, \beta_j, R_j) \sigma_{\rm H,Xq+} Y_{\rm \gamma,H} + 
  N_{\rm He}(\lambda_j, \beta_j, R_j) \sigma_{\rm He,Xq+} Y_{\rm \gamma,He}
  \right] ds
\end{equation}

\noindent where $F_{\rm Xq+}(R)$ is the ion flux as a function of the distance from the sun ($R$) and is assumed to decrease as the inverse square of $R$, $N_{\rm H}(\lambda, \beta, R)$ is the H neutral density at the position $(\lambda, \beta, R)$ in ecliptic coordinates, $\sigma_{\rm H,Xq+}$ is the velocity-dependent cross section for the charge exchange interaction, $Y_{\rm \gamma,H}$ is the photon yield for the spectral line $\gamma$ induced by charge exchange on H, and $ds$ is the step size along the LOS which is variable, being small near the Earth and increasing with distance from the Earth\cite{Koutroumpa2007}. The same definitions hold for He neutrals. The H and He neutral simulations are described in detail in\cite{Koutroumpa2006}, based on models developed in\cite{Lallement1984,Dalaudier1984}. Extended Data Fig.~3 illustrates the sum and shows scaled H and He neutral densities as a function of radial distance from the Sun and the corresponding time of flight for a typical solar wind speed of 400~km/s.

Ideally, we would perform the sum using the instantaneous ion flux at each position along the LOS up to 100 Astronomical Units (AU). However, acquiring those data would require a large flotilla of spacecraft that are not available. We use data from the Solar Wind Ion Composition Spectrometer\cite{SWICS_ref} (SWICS) on the Advanced Composition Explorer (ACE) which is positioned at the L1 Lagrange point between the Earth and Sun. To estimate the ion flux at each $R_j$ for a given HaloSat observation, we look back in time to find which `local packet' of solar wind propagated to the distance $R_j$ at the time of the observation. We perform this calculation using the measured time history of the solar wind speed. To help account for the fact that the solar wind is not uniform across longitude, we use SWICS data acquired when ACE was positioned in the same ecliptic quarter of the sky as the Earth during each HaloSat pointing. Data on the solar wind out of the ecliptic plane is not available for the HaloSat observations. Instead, we used the global average heliolatitude profile of the solar wind speed derived by two different methods, interplanetary scintillation and Lyman-alpha mapping of neutral H with SOHO/SWAN\cite{Koutroumpa2019}, to calculate the solar wind speed along each LOS for each target. We find that the solar wind speed is below 650~km/s for all targets except one and that the slow solar wind (speed below 550~km/s) dominates the most emissive parts of the LOS ($R < 3-5 \rm \, AU$).

The SWICS instrument on ACE prior to 23 August 2011 was capable of measuring the ionic charge composition of the solar wind providing absolute densities and velocities (thus fluxes) as well as abundance ratios. However, a subsequent radiation and age-induced hardware anomaly reduced the capabilities of the instrument and direct measurements of the absolute O$^{7+}$ and O$^{8+}$ fluxes were not available during the period of the HaloSat observations. To estimate these ion fluxes we use an empirical model based on the $\rm O^{7+}/O^{6+}$ flux ratio, which is currently available for SWICS, that we calibrated using SWICS data from before the anomaly. The relations are

\begin{gather}
F_{\rm O7+} = 0.38582 - 0.40041 \exp[-2.7107 (\rm O^{7+}/O^{6+})] \\
F_{\rm O8+} = 0.15574 - 0.1637 \exp[-((\rm O^{7+}/O^{6+}) - 0.00094032)/1.3604]
\end{gather}

\noindent{\bf Model Fitting.} Two different fitting techniques were used to account for the large intrinsic EM variations. The first is use of a `patchiness parameter' originally introduced by\cite{Savage1990} that has been used previously in the literature for similar anlyses\cite{Savage2009,Qu2019}. The intrinsic scatter is modeled as the patchiness parameter ($\sigma_p$), which is combined with the measurement uncertainty ($\sigma_m$) to obtain the uncertainty adopted in the model fitting  $\sigma_f^2= \sigma_p^2+ \sigma_m^2$. The patchiness parameter is then adjusted to give a reduced $\chi^2$ value near unity. We adjusted the patchiness parameter using the final fit, for the sum of the emprical disc and abiabatic polytrope, then fixed $\sigma_p = 3.4 \times 10^{-3} \rm \, cm^{-6} \, pc$ for all of the model fits to allow comparison of the quality of fits between different models. The second fitting method is use of the Huber loss function which is commonly used in robust regression\cite{Ivezi2014}. The Huber loss function includes all of the data points with no addition to the measurement uncertainties, but decreases the weighting for data points with large devations from the model. The margin between inlier and outlier residuals was set at 2. The two fitting methods produce consistent results as shown in Extended Data Fig.~4. The statistical errors are quoted at the $1\sigma$ confidence level. We quote the results from the patchiness parameter method in the main body of the paper. To provide an independent means to evaluating the goodness of fit of the different models, we also calculate the $\chi^2$ and number of degrees of freedom (DoF) after removing 10\% and 20\% of the largest outliers with no addition to the measurement uncertainties. The large reduced $\chi^2$ values arise because our model is smooth while the actual Galaxy is not.

\section*{Data and Code Availability}
The OMNI data are available at \url{https://omniweb.gsfc.nasa.gov/} and ACE data at \url{http://www.srl.caltech.edu/ACE/ASC/level2/lvl2DATA\_SWICS\_SWIMS.html}. The first year of HaloSat data are available at NASA's HEASARC. The additional HaloSat data analysed in this study are available from the corresponding author on request. The computer code used to analyse the data in this study is available from the corresponding author on request. 


\clearpage

\begin{center}
\includegraphics[width=5in]{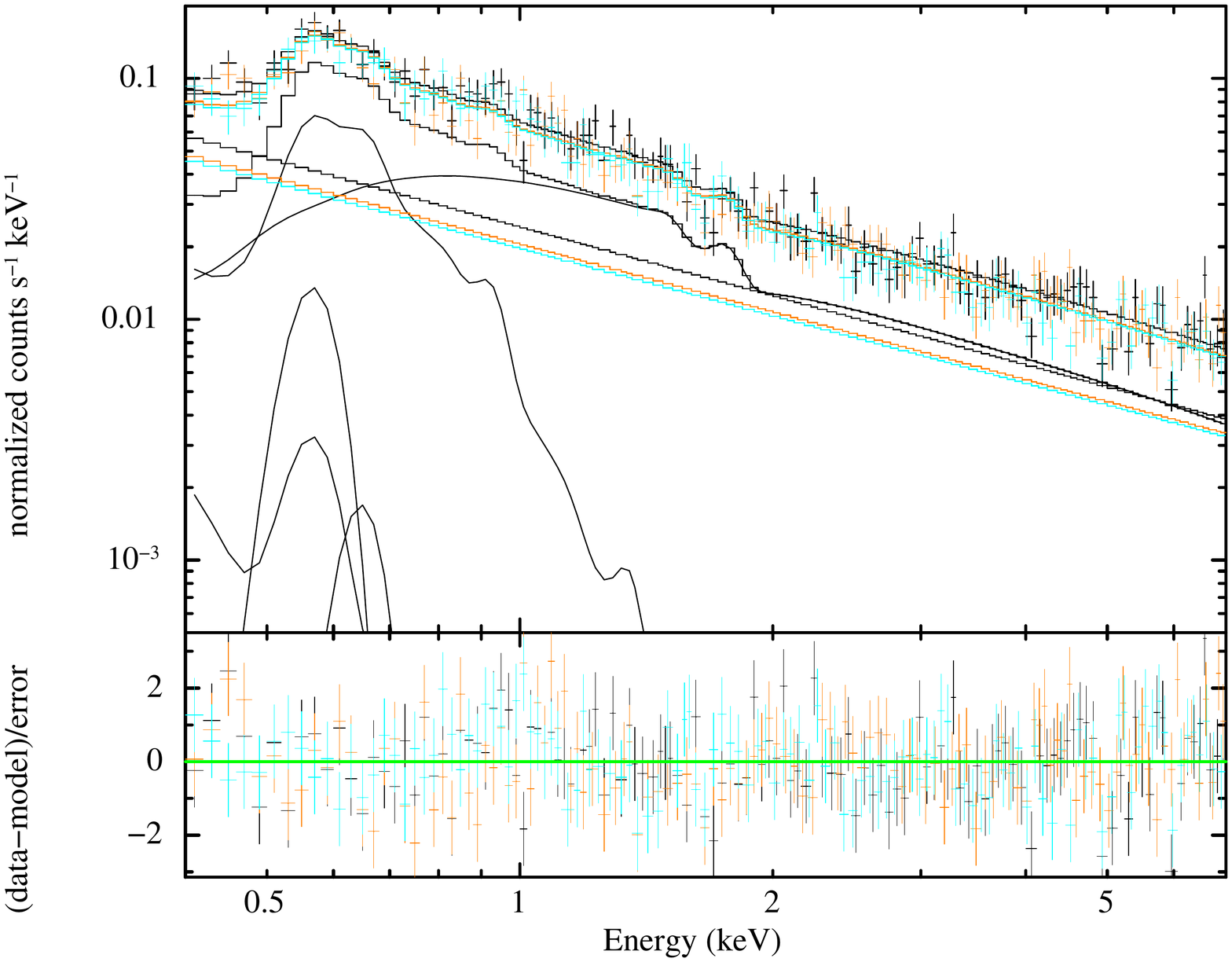} 
\end{center}

\noindent{\bf Extended Data Fig.~1 | HaloSat X-ray spectra.} X-ray spectra of the HaloSat field at $(l = 122.616^{\circ}, b = -55.418^{\circ})$. Data from the three detectors are shown, detector 14 in black, 54 in orange, and 38 in blue. Errors indicate 1-$\sigma$ confidence intervals. The average exposure per detector is 28~ks. The best fitted model and the powerlaw used to model the instrumental background is shown in the same colour for each detector. The model components are shown as black lines. At 0.6~keV, the highest is the sum of the astrophysical components (all components except the instrumental background), the thermal plasma for the halo, the instrumental background, the cosmic X-ray background, the O{\sc vii} oxygen line, the thermal plasma for local hot bubble, and the O{\sc viii} oxygen line. 

\bigskip
\bigskip
\bigskip


\noindent{\bf Extended Data Figure~2 | HaloSat instrumental background model}. The photon index of the instrumental background for each detector is calculated from the count rate in the 3--7~keV band using the equation below with the slope and intercept values in the table.

\noindent Photon index = slope*(hard rate - hr0) + intercept, where hr0 = 0.05 counts~s$^{-1}$.

\begin{center}
\begin{tabular}{c|c|c} \hline
\textbf{Detector} & \textbf{Slope ($s^{-1}$)}    & \textbf{Intercept} \\
  14              &  -4.404           &  0.924 \\
  54              &  -6.138           &  0.890 \\
  38              &  -5.688           &  0.877 \\ \hline
\end{tabular}
\end{center}

\clearpage
\begin{center}
\includegraphics[width=6in]{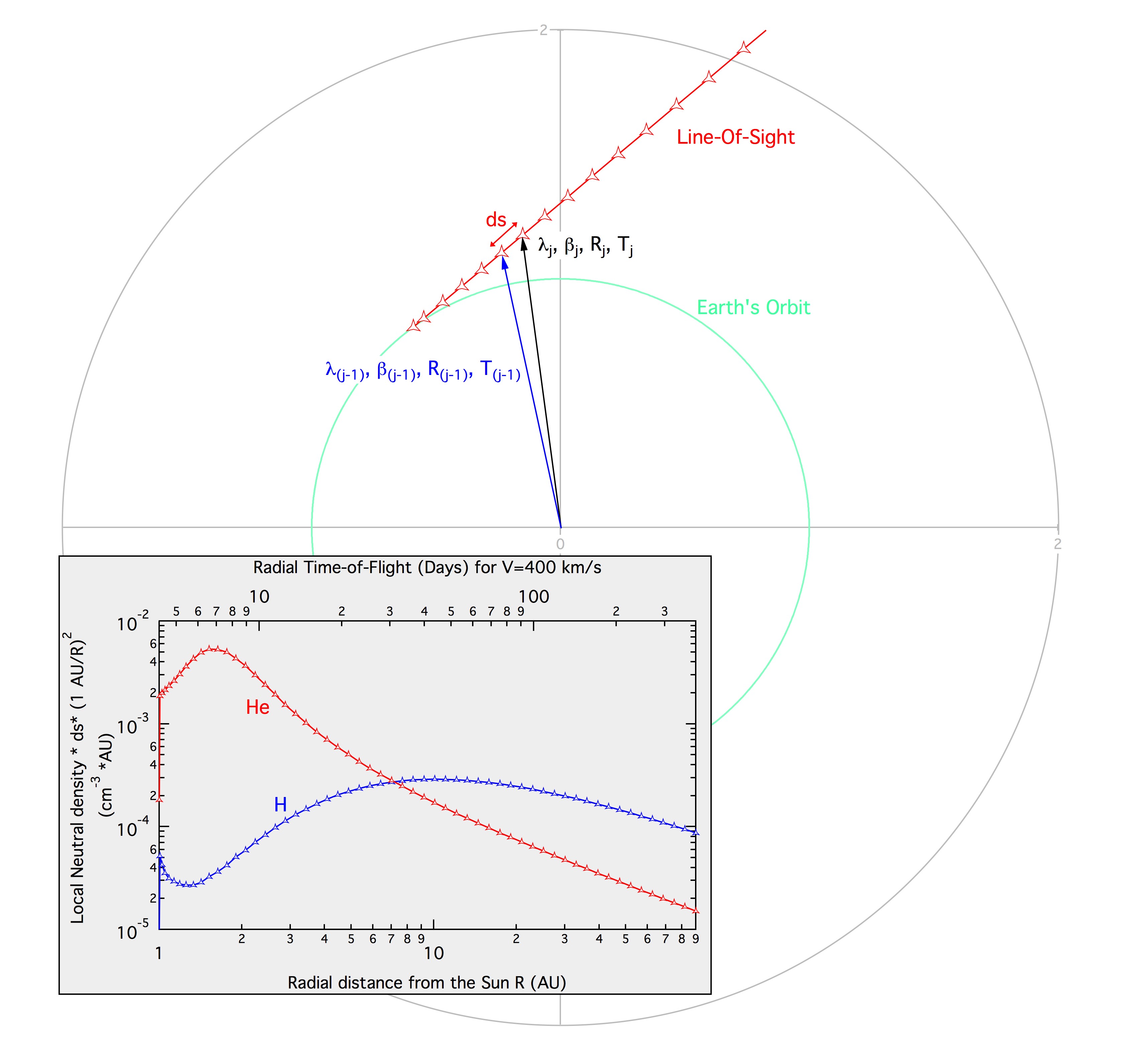}
\end{center}

\noindent{\bf Extended Data Fig.~2 | Heliospheric solar wind charge exchange.} Illustration of the sum used to calculate the line flux due to heliospheric solar wind charge exchange (SWCX) along a line of sight. The Sun is at the origin and the figure is limited to the first 2~AU. Ecliptic longitude is zero along the horizontal axis to the right and increases anti-clockwise. Each triangle represents a term in the sum at the radial distance form the Sun ($R_j$) with the solar wind intensity evaluated for the time of flight $T_j$ from a radius of 1~AU. The inset shows the H and He neutral densities scaled by the step size ($ds$) and by (${\rm 1 AU}/R_j)^2$ as a function of radial distance from the Sun for the full radial scale of the simulations. The density profiles were calculated for the Earth at $126.56^{\circ}$ ecliptic longitude and a look direction of ($\lambda = 39.88^{\circ}, \beta = -7.64^{\circ})$ that crosses the He-focusing cone and the H-ionization cavity resulting in a high oxygen line flux and a high He/H neutral density ratio.


\clearpage
\noindent{\bf Extended Data Fig.~4 | Density Model Fit Results.} Results of fitting the EM data to various density profiles. The first column specifies the model, model parameters, and goodness of fit statistics. Results from fitting using a patchiness parameter are given in the second column and using a Huber loss function in the third column.  The Fit statistic is either $\chi^2$ with the patchiness parameter included or the value of the Huber loss function.

\begin{center}
\begin{tabular}{lcc} 
\textbf{Parameter}  & \textbf{Patchiness}
                                        & \textbf{Huber} \\
\hline
\multicolumn{3}{l}{Modified beta halo} \\
$n_c$ (cm$^{-3}$)   & 0.0044$\pm$0.0004 & 0.0043$\pm$0.0003  \\
$\beta$             & 0.390$\pm$0.021   & 0.395$\pm$0.015    \\
Fit statistic/DoF   & 98.2/71           & 166.10/71          \\
$\chi^2$/DoF (10\%) & 322.7/63          & 363.6/63           \\
$\chi^2$/DoF (20\%) & 271.1/56          & 342.0/56           \\
\hline
\multicolumn{3}{l}{Adiabatic halo} \\
$\rho_V$ ($10^{-5} \rm \, cm^{-3})$
                    & 9.2$\pm$1.5       & 8.4$\pm$1.0        \\
Fit statistic/DoF   & 247.0/71          & 289.2/71           \\
$\chi^2$/DoF (10\%) & 719.9/63          & 782.8/63           \\
$\chi^2$/DoF (20\%) & 564.7/56          & 714.9/56           \\
\hline
\multicolumn{3}{l}{Exponential disk} \\
$n_0$ (cm$^{-3}$)   & 0.012$\pm$0.007   & 0.008$\pm$0.003    \\
$R_0$ (kpc)         & 5.4$\pm$1.5       & 6.2$\pm$1.1        \\
$z_0$ (kpc)         & 2.8$\pm$1.0       & 3.8$\pm$0.8        \\
Fit statistic/DoF   & 87.7/70           & 151.8/70           \\
$\chi^2$/DoF (10\%) & 267.9/62          & 294.9/62           \\
$\chi^2$/DoF (20\%) & 214.4/55          & 263.1/55           \\
\hline
\multicolumn{3}{l}{Empirical disk} \\
$n_0$ (cm$^{-3}$)   & 0.0144$\pm$0.0007 & 0.0134$\pm$0.0004  \\
$z_0$ (kpc)         & 1.09$\pm$0.08     & 1.2$\pm$0.5        \\
Fit statistic/DoF   & 75.8/71           & 139.2/71           \\
$\chi^2$/DoF (10\%) & 235.3/63          & 247.7/63           \\
$\chi^2$/DoF (20\%) & 196.4/56          & 220.0/56           \\
\hline
\multicolumn{3}{l}{Empirical disk plus adiabatic halo} \\
$\rho_V$ ($10^{-5} \rm \, cm^{-3})$
                    & 4.8$\pm$1.0       & 3.3$\pm$1.0        \\
$n_0$ (cm$^{-3}$)   & 0.0081$\pm$0.0022 & 0.0100$\pm$0.0015  \\
$z_0$ (kpc)         & 1.60$\pm$0.34     & 1.42$\pm$0.16      \\
Fit statistic/DoF   & 71.7/70           & 137.0/70           \\
$\chi^2$/DoF (10\%) & 231.2/62          & 246.0/62           \\
$\chi^2$/DoF (20\%) & 191.3/55          & 218.5/55           \\
\hline
\end{tabular}
\end{center}

\end{document}